\documentclass{W-template}
\usepackage{color}
\begin{document}
\pagestyle{fancy}

\title{Uniform deposition of particles in large scale by drying of binary droplets}

\maketitle

\author{Zechao Jiang$^{\dag}$}
\author{Liyiming Tao$^{\dag}$}
\author{Xiuyuan Yang}
\author{Masao Doi}
\author{Ye Xu*}
\author{Xingkun Man*}

\dedication{$^{\dag}$ Z.Jiang contributed equally to this work with L. Tao.}

\begin{affiliations}
Z. Jiang\\
School of Physics, Beihang University, Beijing, 102206, China

L. Tao\\
School of Mechanical Engineering and Automation, Beihang University, Beijing, 102206, China

X. Yang\\
School of Physics, Beihang University, Beijing, 102206, China

M. Doi\\
School of Physics, Beihang University, Beijing, 102206, China \\
Wenzhou Institute, University of Chinese Academy of Science, Whenzhou, 325000, China

Y. Xu\\
School of Mechanical Engineering and Automation, Beihang University, Beijing, 102206, China\\
Email Address: ye.xu@buaa.edu.cn

X. Man\\
School of Physics, Beihang University, Beijing, 102206, China\\
Peng Huanwu Collaborative Center for Research and Education, Beihang University, Beijing\\
Email Address: manxk@buaa.edu.cn

\end{affiliations}


\keywords{evaporation, coffee-ring effect, binary droplets, Marangoni flow, uniform deposition}

\begin{abstract}

The evaporation of liquid droplets often results in a ring-like deposition pattern of particles, presenting challenges for applications requiring highly uniform patterns. Despite extensive efforts to suppress the coffee ring effect, achieving a uniform particle distribution remains a great challenge due to the complex and non-equilibrium nature of the evaporation process. In this work, we introduce and demonstrate a one-step drying method for binary droplets (water and 2-methoxyethanol) that produces uniform deposition of nano- and micro-particles. By adjusting the initial water volume fraction, we effectively control the interplay between capillary and Marangoni flows, resulting in deposition patterns that vary from coffee ring to uniform and to volcano-like. Through both theoretical and experimental analyses, we determine the conditions necessary for achieving such high uniformity. This approach requires no special substrate treatment, particle modification, or controlled environments, and works for various particles, including silica and polystyrene. Our method provides a robust solution for fabricating uniform patterns that are crucial for many practical applications, ranging from printing to microelectronics to bio-pharmacy.

\end{abstract}


\section{Introduction}

The evaporation of liquid droplets containing nonvolatile solutes typically results in a ring-like deposition pattern. This phenomenon is frequently observed in nature, and is known as the coffee-ring effect (CRE) \cite{deegan_1997_capillary, degennes_2013_capillarity, Wilson_2023_Evaporation}. Nevertheless, in many practical applications, ranging from inkjet printing to biomedical engineering, there is a need to suppress the coffee-ring effect and require uniform deposition patterns \cite{dugas_2005_droplet, sekitani_2009_stretchable, choi_2016_exploiting, deng_2018_surfactantcontrolled, dahiya_2019_largearea, Yang_2020_HighResolution, su_2022_3dprinted, kim_2022_exploiting, chen_2022_perovskite, Fan_2022_Bionic, pal_2023_drying, Fonseca_2023_3D, Wang_2024_Achieving}. Despite considerable efforts devoted to suppress the CRE \cite{Anyfantakis_2015_Manipulating, Mampallil_2018_review, Brutin_2018_Recent, zang_2019_evaporation, Mondal_2023_Physics}, achieving uniform particle distribution through one-step drying of liquid droplets remains unresolved and is a formidable challenge \cite{lohse_2022_fundamental}. This is due to the multitude and complexity of physical and chemical factors associated with particle deposition, including the temperature and relative humidity, the characteristics of the solution, the size and shape of the particles, the surface energy of the substrate, and so on.

Over the past decades, numerous methods have emerged to suppress the CRE. Electrowetting techniques utilize external electric fields to induce radial electroosmotic flows or prevent contact line pinning, thereby promoting uniform particle distribution \cite{kim_2006_control, eral_2011_suppressing}. Adding surfactants can induce Marangoni flows or facilitate complex sol-gel transitions, which counteract outward capillary flow and lead to central particle accumulation \cite{still_2012_surfactantinduced, sempels_2013_autoproduction, talbot_2015_printing, kim_2016_controlled}. Salt is also commonly used to adjust the interfacial properties of the emerging crystal \cite{Shahidzadeh_2015_Salt}. Modifying substrates by altering the wettability, porosity, or surface energy can also control the dynamics of the contact line and effectively suppress the CRE \cite{li_2013_evaporation, Pack_2015_Colloidal, li_2020_evaporating}. Adjusting the size, shape, and interaction between particles influences their movement during evaporation, with ellipsoidal particles, modified particles and amphiphilic microgels showing promise in achieving uniform deposits \cite{Bigioni_2006_Kinetically, yunker_2011_suppression, ge_2017_colloidal, Mayarani_2017_Loosely, Bansal_2018_Suppression, Koshkina_2021_Surface, rey_2022_versatile, Jose_2023_Depletion}. The temperature gradient \cite{Hu_2006_Marangoni, Lama_2020_Modulation} can induce the surface tension gradient along the liquid-vapor (L-V) interface, generating the Marangoni flow that adjusts the internal flow field and particles distribution. Furthermore, changing the speed and distribution of evaporation through ambient humidity or substrate heating can further modulate the spatial distribution of colloids \cite{Fukuda_2013_Profile, zargartalebi_2022_selfassembly}. Other methods include adjusting the pH to influence DLVO interactions \cite{bhardwaj_2010_selfassembly, almilaji_2018_phmodulated} and employing external thermal or vapor sources \cite{Ta_2016_Dynamically, malinowski_2018_dynamic}. Overall, these strategies provide a comprehensive understanding and offer a range of approaches applicable to different systems for suppressing the CRE.

However, the suppression of CRE does not ensure the attainment of uniform deposition of particles. Here, uniform means that the experiment measured root mean square ($\rm Rms$) values of the normalized height $h(r)/D$ is less than $0.1$, where $D$ is the diameter of the solute particles. In fact, reports of uniform deposition achieved solely by the evaporation of liquid droplets are rare. Jaeger et al. demonstrated that adding an optimal amount of dodecanethiol facilitates the formation of a uniform monolayer film in the evaporation of a solution containing monodisperse ($<5\%$) dodecanethiol-ligated gold nanocrystals in toluene \cite{Bigioni_2006_Kinetically}. Satapathy et al. successfully achieved two-dimensional particle deposits by controlling the evaporation of soft pNIPAM microgel particles through concentration tuning \cite{Mayarani_2017_Loosely}. Vogel et al. achieved homogeneous solute patterns using surface-modified particle dispersions \cite{rey_2022_versatile}. By employing high-molecular-weight surface-active polymers that physisorb onto particle surfaces, they enhance steric stabilization and mitigate edge pinning during droplet evaporation. Similarly, Sanati-Nezhad et al. obtained highly ordered particle deposits on microscale surface areas by placing droplets on near-neutral-wet shadow molds attached to hydrophilic substrates \cite{zargartalebi_2022_selfassembly}. While these strategies have advanced the field significantly, they often rely on specialized treatments or specific conditions, limiting their reproducibility and scalability for large-scale manufacturing. The pursuit of uniformly depositing particles, crucial for the performance of modern electronic devices, calls for low-cost, one-step and universally applicable methods \cite{diao_2014_morphology, zhang_2018_ultrasmooth, zhang_2020_recent}.

Deposition patterns during droplet evaporation are mainly influenced by the internal evaporation-induced flow field. Consequently, most strategies for pattern control are centered on manipulating capillary and Marangoni flows. Hu and Larson demonstrated that a surface tension gradient, higher in the center and lower at the edges of the droplet, generates an inward Marangoni flow, inhibiting the coffee ring effect and resulting in mountain-like deposition patterns \cite{Hu_2006_Marangoni}. However, this surface tension gradient, produced by the temperature gradient as a result of the latent heat of evaporation, makes it challenging to control the magnitude of the Marangoni flow. Specifically, a significant Marangoni flow can be achieved in Octane droplets, whereas in water droplet systems, the Marangoni flow is too weak to inhibit coffee rings. The evaporation of binary droplets offers promising solutions by allowing for the adjustment of the Marangoni flow through variations in concentration and type of the binary mixture liquids \cite{kim_2006_direct, lim_2008_selforganization, liu_2016_line, zhong_2016_flow, hu_2017_black, shi_2019_drying, hu_2020_a, gao_2020_printable, pahlavan_2021_evaporation}. Besides, related binary droplets studies found rich contact line motion behavior and internal flow fields \cite{Wang_2022_Wetting}, including Marangoni contraction \cite{Karpitschka_2017_Marangoni}, Marangoni instability \cite{Keiser_2017_Marangoni}, and phase segregation \cite{Li_2018_Evaporation}. However, experiments \cite{kim_2006_direct, lim_2008_selforganization, liu_2016_line, zhong_2016_flow, hu_2017_black, shi_2019_drying, hu_2020_a, gao_2020_printable, pahlavan_2021_evaporation} showed contradictions that the CRE can be inhibited by both inward and outward Marangoni flow, which implies potential discrepancies in the mechanisms involved. Moreover, systematic discussion about how to utilize the evaporation of binary droplets to achieve uniform deposition is still lacking and remains highly challenging in this complex non-equilibrium system.

Here we introduce and demonstrate the achievement of uniform deposition of nano/micro-particles using the evaporation of binary droplets. It is shown that the synergetic effects of capillary flow and Marangoni flow on the convection of suspended particles can be regulated by adjusting the initial volume fraction ratio of the two solvent components in the droplet. When the velocity magnitude and the particle-convection time of the two flows are comparable, uniform deposition of particles can be achieved. We determine, both theoretically and experimentally, the conditions of generating comparable capillary and Marangoni flows in evaporating binary droplets. Moreover, it is shown that as long as such conditions are satisfied, particles with various types and initial concentration can be deposited uniformly using the evaporation of binary droplets. We provide a robust methodology for fabricating uniform deposition of particles using binary droplets without the need for any specialized treatment of the liquid, particle, substrate or ambient environment.

\section{Results and Discussion}

\subsection{Characterization of the Deposition}

\begin{figure}[!tbh]
  \centering
  \includegraphics[width=0.8\linewidth]{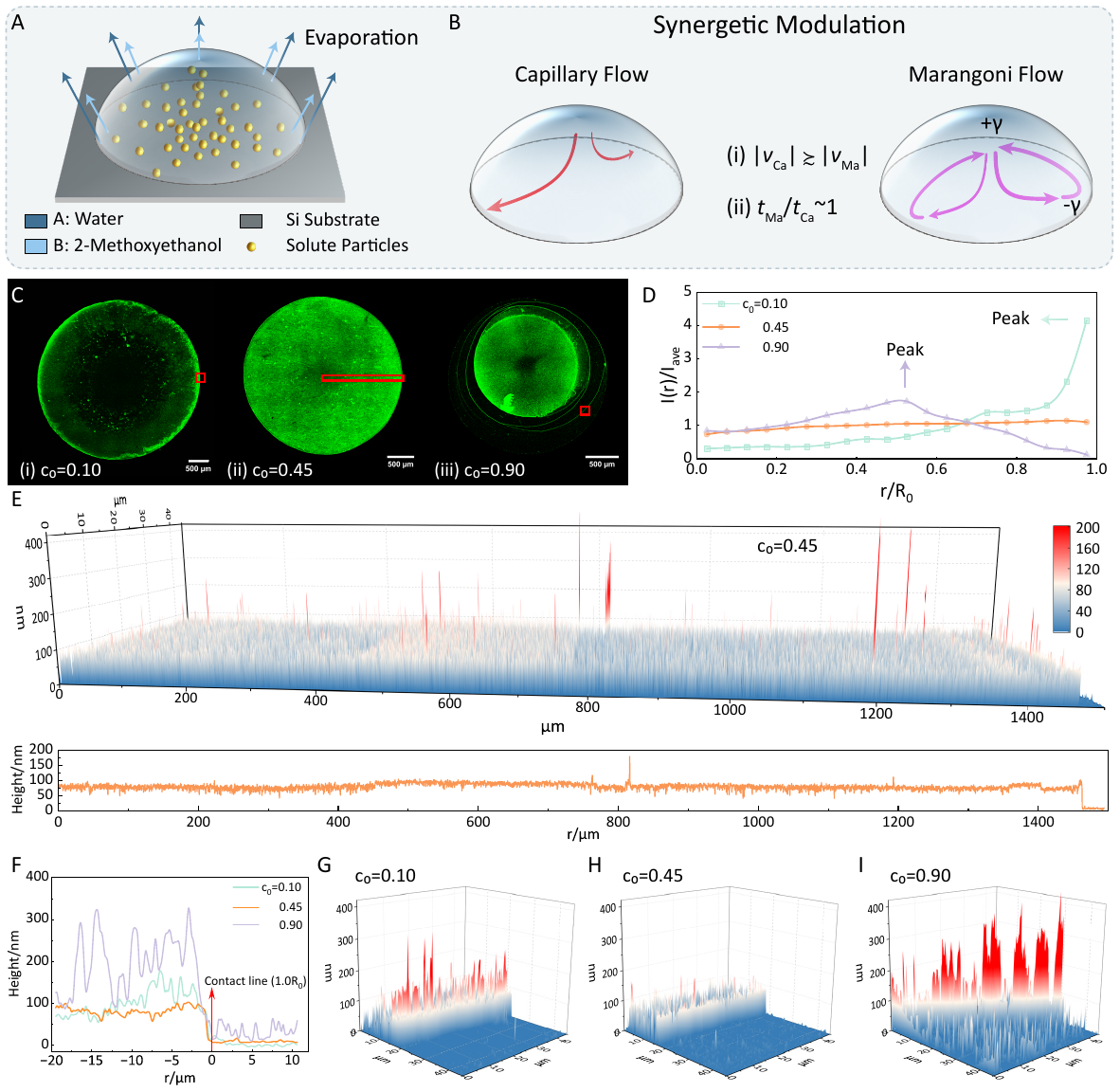}
  \caption{Design strategy and characterization of deposition patterns of drying binary droplets. (A) Schematic of a binary droplet of water and 2-methoxyethanol (2-ME) ladened with PS particles (diameter $100{\rm nm}$) evaporate on Si substrates. (B) Harnessing the capillary and Marangoni flow of this evaporating droplet to obtain uniform distribution of deposited particles. Two conditions are required: (i) the velocity of the capillary flow and the Marangoni flow should be comparable $|v_{\rm Ca} | \geq |v_{\rm Ma} |$; (ii) the convection duration of the two flows on the particles has to be close $t_{\rm Ma}/t_{\rm Ca} \sim 1$. (C) The fluorescence microscope images of deposition patterns. The initial volume fraction of water $c_0$ in the droplet is (i) $0.10$, (ii) $0.45$ and (iii) $0.90$, respectively. (D) Normalized light intensity $I(r)/I_{\rm ave}$ of deposited particles along the normalized radius $r/R_0$ are plotted for each of the images in (C), respectively. $R_0$ is the initial contact radius. $I(r)=(1/2\pi) \int_0^{2\pi} I(r,\theta)d\theta$ is the local light intensity and $I_{\rm ave}$ is the average of $I(r)$. (E-I) Three-dimensional (3D) views and 2D side views profiling of the distribution of deposited particles using atomic force microscopy (AFM) topography images. (E) is the height morphology of the deposition pattern in the red-frame in C(ii) that is a region of $45{\rm \mu m}\times 1400{\rm \mu m}$ from the center to the edge. (F) is the comparison of 2D side view of the height of deposition patterns in the red-framed regions in C(i)-C(iii) located at the droplet edge ($\sim 1.0R_0$). (G-I) are the corresponding 3D height morphologies ($45{\rm \mu m}\times 45{\rm \mu m}$) in (F). The color bar represents the morphology height, applicable to (E, G-I).}
  \label{fig1}
\end{figure}

Identifying the conditions for uniform deposition of particles is challenging through trial-and-error, given the complex non-equilibrium nature of the evaporation of binary droplets. To tackle this challenge, we conduct a study that combines both theory and experiment. We obtain uniformly deposited particles through the drying of binary droplet, ladened with fluorescent-labeled polystyrene (PS) particles, with a mixture of water and 2-methoxyethanol (2-ME) on the surface of silicon substrates in environmental conditions (Fig.\ref{fig1}A and see Methods and SI Appendix for details). The initial volume $V_0$ of the used droplet is $1.5{\rm\mu L}$, the diameter of PS particles is $100 {\rm nm}$, and the initial concentration is unit $1\times$ ($0.0005\% {\rm vol}$). Fig\ref{fig1}.C is the fluorescence microscope images of the deposited PS particles. When the initial volume fraction of water $c_0$ is $0.10$, a coffee-ring of PS particles are formed. When $c_0$ is $0.45$, PS particles are deposited uniformly on the substrate. As $c_0$ increases to $0.90$, the deposition pattern becomes a volcano-like ring pattern.

To quantify the uniformity of deposited particles, we analyze the fluorescent signals of the intensity distribution of the three deposition patterns shown in Fig.\ref{fig1}C. Fig.\ref{fig1}D represents the normalized fluorescent intensity distribution $I(r)/I_{\rm ave}$ of the deposition patterns, where $I(r)=(1/2\pi) \int_0^{2\pi} I(r,\theta) d\theta$ is the local fluorescent intensity on a ring with a radius $r$ from the center of the pattern, and $I_{\rm ave}$ is the averaged value of $I(r)$. The fluorescent intensity distribution reveals that, for the case corresponding to $c_0=0.10$ (green data points), there is a peak in light intensity near the initial contact line $R_0$, indicating that most PS particles are deposited near $R_0$ forming a coffee-ring. When the initial concentration is $0.90$ (purple data points), the peak of the light intensity is near $0.5R_0$, indicating that most PS particles are deposited between the initial boundary and the center of the droplet, forming a volcano-like pattern. In significant contrast, when the initial concentration is $0.45$ (orange data points), the light intensity remains nearly constant throughout the entire range, suggesting a uniform particle number density in the deposition region.

To further quantify the distribution of deposited particles, we use Atomic Force Microscopy (AFM) to measure the height profiles of the three deposition patterns shown in Fig.\ref{fig1}C. The AFM measurements were performed on the red-framed regions indicated in Fig.\ref{fig1}C. For the case of $c_0=0.10$, the AFM measurement is conducted in the area near the edge of the deposited pattern and shows a height profile with peaks around $250{\rm nm}$ (Fig.\ref{fig1}G). This clearly indicates a vertical stacking of deposited particles because the diameter of PS particles that we used is $100{\rm nm}$. Similarly, the height profile of the case of $c_0=0.90$ also indicates vertical stacking of deposited particles (Fig.\ref{fig1}I). By contrast, the height distribution for the droplet with $c_0=0.45$ exhibits remarkable uniformity around $100{\rm nm}$, as shown in the stitched AFM image covering a large region from the edge all the way to the center of the deposited pattern (Fig.\ref{fig1}E), indicating consistent height distribution of the deposited patterns at different locations. Additionally, statistical analysis of the AFM height distribution yields an average roughness ${\rm Ra}=6.0\pm 0.8 {\rm nm}$ and a root mean square roughness ${\rm Rms}=7.9\pm 1.2 {\rm nm}$ (see SI Appendix Fig.S9). These results collectively indicate the obtained pattern is a monolayer deposition of PS sphere particles.

\subsection{Conditions for uniform Deposition}

\begin{figure}[!tbh]
       \centering
       \includegraphics[width=0.8\linewidth]{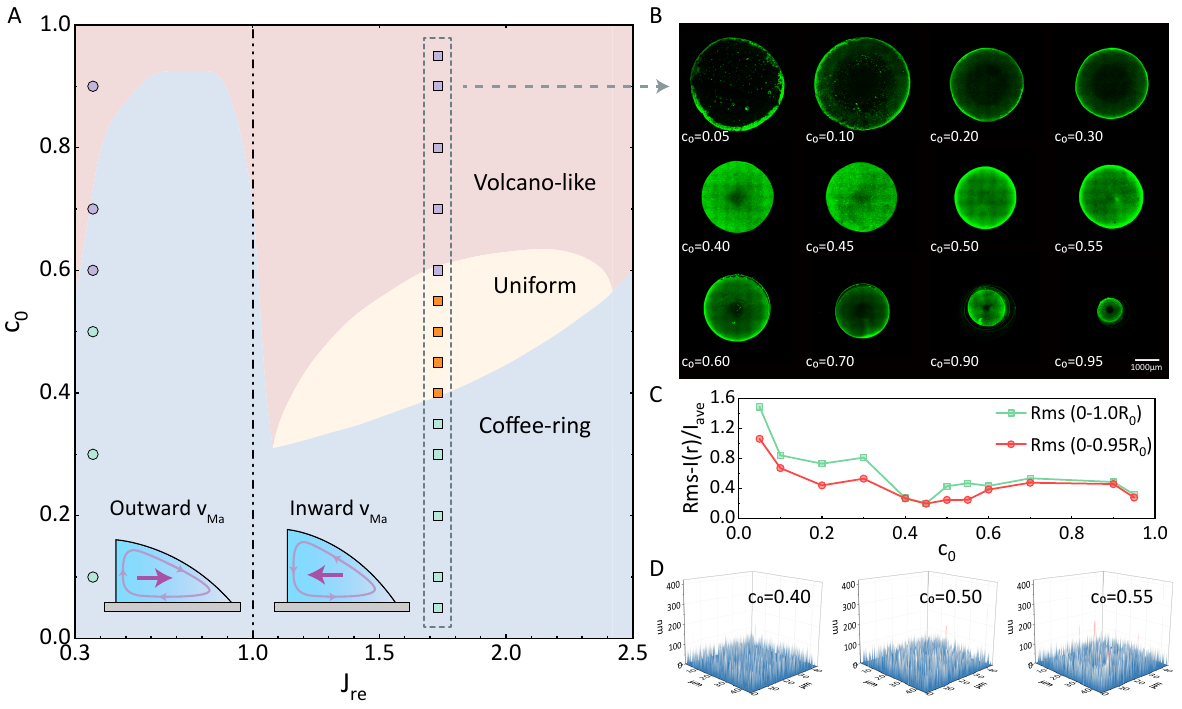}
       \caption{Phase diagram of deposition pattern of drying binary droplets. (A) Calculated phase diagram of deposition pattern in terms of the initial volume fraction of solvent A $c_0$ and the evaporation rate ratio of the two solvents $J_{\rm re}$. All circle and square data points are experimental results, where circle represents water/ethanol droplets while square is for water/2-ME droplets. Results show three typical deposition patterns that are coffee-ring (blue region, green circle/square), volcano-like (pink region, purple square) and uniform (yellow region, orange circle/square). Inset: Schematic of the direction of the Marangoni flow $v_{\rm Ma}$ for $J_{\rm re}<1.0$ (outward) and $J_{\rm re}>1.0$ (inward), respectively. (B, C) The fluorescence microscope images and the root mean square ($\rm Rms$) of normalized light intensity $I(r)/I_{\rm ave}$ of deposition patterns for experimental results in the dashed rectangle in (A), where $c_0$ increases from $0.05$ to $0.95$. Green line and red line are the $\rm Rms$ within the range of $0-1.0R_0$ and $0-0.95R_0$ of the deposition area, respectively. (D) AFM topography images of the distribution of deposited particles for $c_0=0.40$, $0.50$ and $0.55$ cases shown in (B). The color represents the morphology height, same to that in Fig.1E. For the parameters in (A), the value of key parameter $\gamma_{\rm re}$ (the surface tension ratio) is $2.6$. All other parameters for calculations can be found in SI.}
       \label{fig2}
\end{figure}

To determine the experimental conditions for achieving uniform deposition pattern of particles, we theoretically calculate the phase diagram of the deposition pattern of drying binary droplets containing two (A and B) mixture solvents. The model, based on the Onsager variational principle \cite{doi_2013_soft,doi_2015_onsager,man_2016_ring,jiang_2020_the}, captures the time evolution of droplet shape and internal flow fields during evaporation, and gives the number density distribution of solute particles deposited on the substrate (Theoretical analysis part in the SI Appendix provides the detailed derivation process). Fig.\ref{fig2}A is the theoretical phase diagram of deposition patterns in terms of the initial volume fraction of the A-component, $c_0$, and the initial evaporation rate ratio of the two components, $J_{\rm re}=J_{\rm A}/J_{\rm B}$. Noticing that the evaporation rate of the binary droplet is $J(r,t)=c_{\rm A}(r,t) J_{\rm A} +c_{\rm B}(r, t) J_{\rm B} $ that changes during evaporation, where $c_{\rm A}(r,t)$ and $c_{\rm B}(r,t)$ are the height-averaged volume fractions at position $r$ and time $t$ for $\rm A$ and $\rm B$ components, respectively. In our theoretical calculations, we fixed the surface tension of the A-component to be larger than it of the B-component, i.e., $\gamma_{\rm re}=\gamma_{\rm A}/\gamma_{\rm B}>1$. Our results can be easily extended to situations where $\gamma_{\rm re}<1$ due to the inherent symmetry in the model.

\begin{figure}[!tbh]
       \centering
       \includegraphics[width=0.8\linewidth]{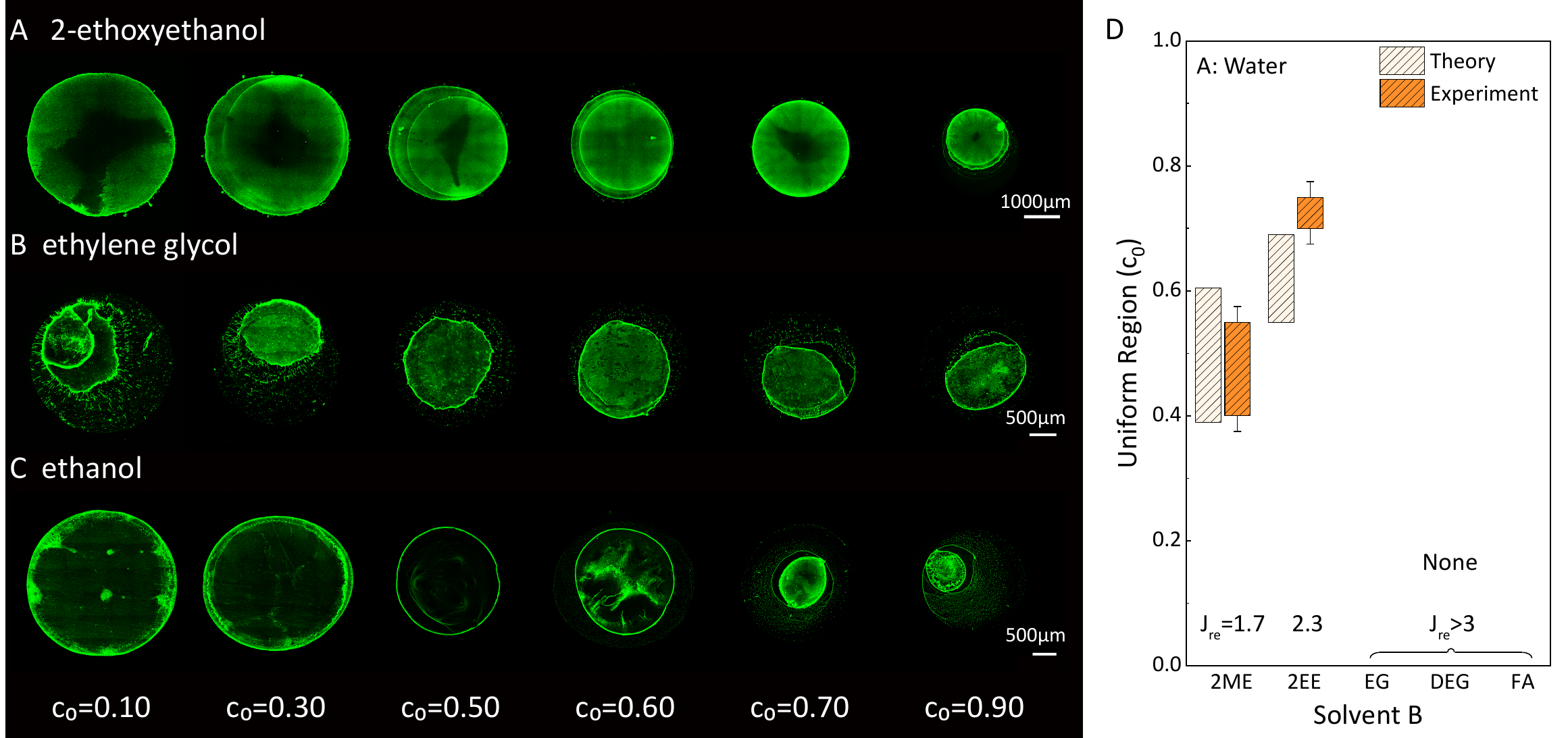}
       \caption{Validation of different binary droplet systems. (A-C) Experimental phase diagram of (A) water/2-ethoxyethanol (2-EE), (B) water/ethylene glycol (EG) and (C) water/ethanol systems on the silicon substrates. (D) The uniform deposition region of the initial volume fraction $c_0$ for different binary droplets, where yellow part is theoretical calculation and orange part is experimental results.}
       \label{fig3}
\end{figure}

The theoretical calculations indicate that when $J_{\rm re}<1$, i.e., the evaporation rate of the solvent with higher surface tension is slower than that of the solvent with lower surface tension ($\gamma_{\rm A}>\gamma_{\rm B}$ and $J_{\rm A}<J_{\rm B}$), only ringlike patterns such as coffee-ring and volcano-like are obtained. Conversely, when $J_{\rm re}>1$ corresponding to the two solvents having $\gamma_{\rm A}>\gamma_{\rm B}$ and $J_{\rm A}>J_{\rm B}$, the deposition pattern transitions from coffee-ring to uniform and then to volcano-like as the initial volume fraction $c_0$ of A-component increases from $0$ to $1.0$. Such theoretical predictions are tested by two experimental systems. The first system is water/ethanol, where water has a higher surface tension than ethanol, but its evaporation rate is slower. Six different droplets with varying initial water volume fraction were tested experimentally (see Fig.\ref{fig3}C and SI Appendix Fig.S4), resulting in coffee-ring (green circle data points) and volcano-like patterns (purple circle data points), aligning well with the theoretical calculations. The second system is water/2-methoxyethanol (2-ME). At room temperature ($20^{\circ}{\rm C}$) and $1{\rm atm}$, water has a surface tension of $\gamma_{\rm W}=72.8 {\rm mN/m}$ and a volume evaporation rate of $J_{\rm W}=(23.2\pm 0.4)\times 10^{-10} {\rm L/(cm^2\cdot s)}$, while 2-ME has a surface tension of $\gamma_{\rm 2-ME}=27.6{\rm mN/m}$ and an evaporation rate of $J_{\rm 2-ME}=(13.4\pm 0.8)\times 10^{-10} {\rm L/(cm^2\cdot s)}$ (see SI Appendix Table S1). Thus, this system has $\gamma_{\rm re}=2.6$ and $J_{\rm re}=1.73$, corresponding to the cases of $\gamma_{\rm A}>\gamma_{\rm B}$ and $J_{\rm A}>J_{\rm B}$. We tested fourteen droplets with different initial water volume fraction ($c_0$ ranging from $0.05$ to $0.95$). Fig.\ref{fig2}B shows the corresponding deposition patterns in fluorescence microscope images, while Fig.\ref{fig2}C presents the $\rm Rms$ of the normalized fluorescent intensity $I(r)/I_{\rm ave}$ for all patterns. The red and green data points correspond to the area of $0-0.95R_0$ and $0-1.0R_0$, respectively. It is evident that the Rms-light intensity for deposition patterns of droplets with $c_0=0.05-0.35$ (green square data points) and $c_0=0.60-0.95$ (purple square data points) is significantly higher than that for patterns obtained at $c_0=0.40-0.55$ (orange square data points). Such experimental results align well with the theoretical calculations. Besides, Fig.\ref{fig2}D represents the measured AFM height distributions corresponding to the uniform deposition $c_0=0.40-0.55$ in Fig.\ref{fig2}C, and the $\rm Rms$ values of the normalized height distribution is less than $0.1$ ($10nm/100nm$, see SI Appendix Fig.S10). Combining the fluorescence microscope images with the AFM results conclusively demonstrates the successful achievement of uniform deposition patterns of PS particles.

Fig.\ref{fig2}A outlines the experimental conditions that are necessary for achieving uniform deposition of particles. 1) a binary droplet of $J_{\rm re}>1$ and $\gamma_{\rm re}>1$, i.e., one solvent having higher surface tension and a faster evaporation rate than the other one. 2) the initial volume fraction of the faster volatile solvent ($c_0$) should fall within a certain range that depends on the value of $J_{\rm re}$. Such range for $c_0$ has the optimal range maximized when $J_{\rm re}$ is around $1.7$. Deviation from this value of $J_{\rm re}$ gradually reduces the permissible range of $c_0$. 3) the evaporation rate ratio of the two solvents ($J_{\rm re}$) also has a limited range. It is shown that uniform deposition of particles is not observed when $J_{\rm re}$ is below $1.1$ or above $2.4$. 

We validated the predicted conditions through four additional binary droplet systems that satisfy $J_{\rm re}>1$ and $\gamma_{\rm re}>1$, including water and 2-ethoxyethanol (2-EE), water and ethylene glycol (EG), water and diethylene glycol (DEG), and water and formamide (FA) systems (see SI Appendix Table S1 for solvent properties). Fig.\ref{fig3}D shows the range of $c_0$ values for achieving uniform deposition of particles in these systems. Theoretical calculations (yellow boxes) and experimental measurements (orange boxes) indicate that the water/2ME system ($J_{\rm re}=1.7$) exhibits the widest range for uniform deposition, while the water/2EE system ($J_{\rm re}=2.3$) (see Fig.\ref{fig3}A) has a range but is smaller than that of the water/2ME system. Meanwhile, the other three systems, with $J_{\rm re}>3$, do not exhibit uniform deposition of particles. For instance, in the water/EG system, only a coffee-ring pattern was observed (see  Fig.\ref{fig3}B).

\subsection{Mechanistic Understanding}
We further elucidate the underlying mechanism of those three conditions in determining the final deposition pattern by analyzing the internal fluid flow in an evaporating binary droplet. As the droplet evaporates, outward capillary flow transports particles from the droplet center to its edge. While such particle transport is necessary to obtain uniform deposition pattern, it is crucial to reduce the number of convected particles to avoid the formation of coffee-ring pattern. To address this, we employ binary droplet systems with $J_{\rm re}>1$ and $\gamma_{\rm re}>1$. These droplets generate a surface tension gradient from the edge towards the center during the evaporation process, generating an inward height-averaged Marangoni flow, i.e., from edge to center. With an appropriately tuned magnitude, the Marangoni flow is expected to balance the capillary flow in convection of particles, thereby leading to the formation of uniform deposition of particles.

\begin{figure}[!tbh]
       \centering
       \includegraphics[width=0.8\linewidth]{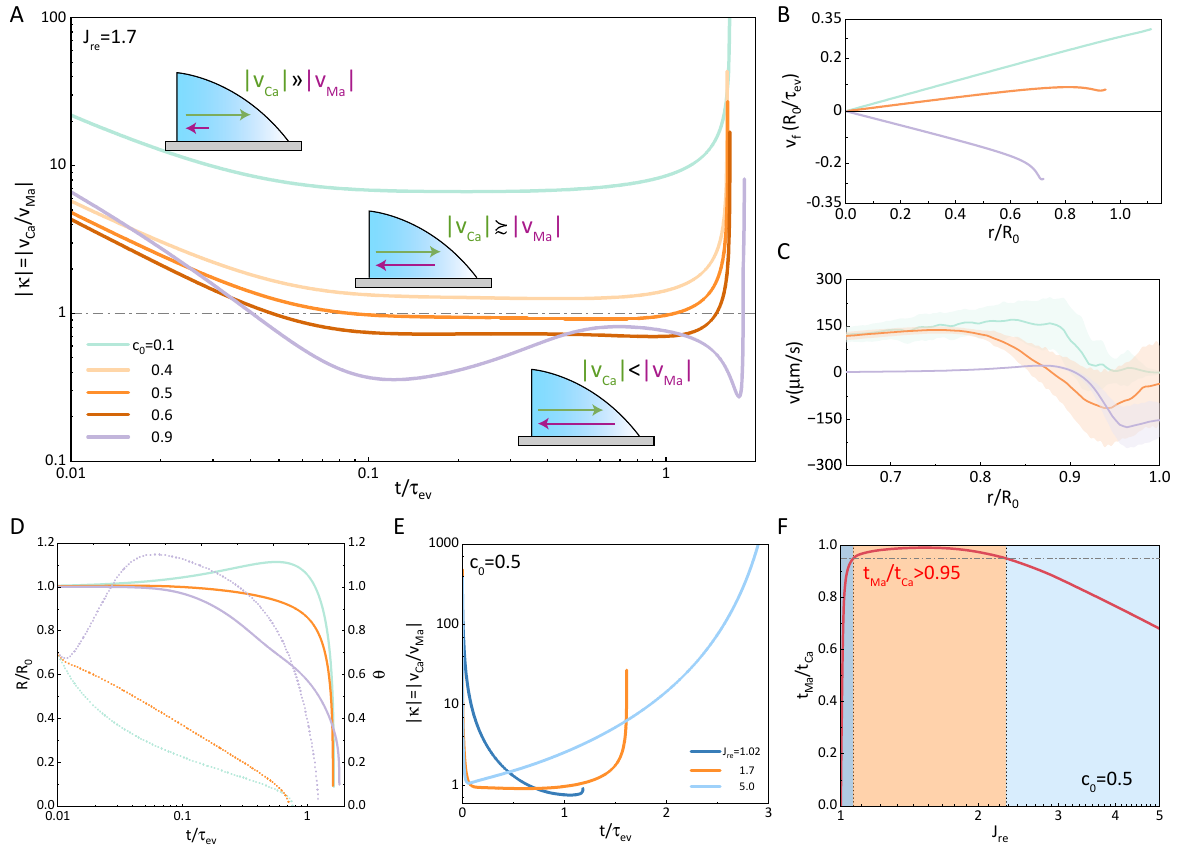}
       \caption{The mechanism underlying the uniform distribution of deposited particles by tailoring the fluid flow of an evaporating binary droplet. (A-D) The explanation for the synergetic modulation strategy (i): the strength of the capillary velocity and the Marangoni velocity should satisfy the condition $|v_{\rm Ca}| \geq |v_{\rm Ma}|$. (A) Theoretical calculation of $|\kappa|=|v_{\rm Ca}/v_{\rm Ma} |$ with normalized time $t/\tau_{\rm ev}$ for various values of $c_0$. Inset: Schematic of the direction and strength of the capillary flow $v_{\rm Ca}$ and the Marangoni flow $v_{\rm Ma}$ for $c_0=0.10$, $0.50$, $0.90$, respectively. (B) Theoretically calculated height-average flow velocity $v_f$ at time $t=0.5\tau_{\rm ev}$ along the normalized radius $r/R_0$ for three droplets characterized by different values of $c_0$. (C) Experimentally measured averaged particle velocity at time $t=0.5\tau_{\rm ev}$ of drying droplets corresponds to those used in (B), with the standard deviation indicated in a lighter shade. (D) Theoretical calculation of normalized contact radius $R/R_0$ (solid lines) and contact angle (dotted lines) with normalized time $t/\tau_{\rm ev}$ for $c_0=0.10$, $0.50$, $0.90$, respectively. The color scale in (A) is also fit for (B-D). (E, F) The explanation and proof for the synergetic modulation strategy (ii): the convection time of the two fluids on the solutes should be the same, i.e., $t_{\rm Ma}/t_{\rm Ca}\sim 1$. (E) Theoretical calculation of $|\kappa|=|v_{\rm Ca}/v_{\rm Ma}|$ with normalized time $t/\tau_{\rm ev}$ for different evaporation rate ratio $J_{\rm re}$. (F) The dependence of $t_{\rm Ma}/t_{\rm Ca}$ on $J_{\rm re}$ for a droplet with $c_0=0.50$, where the dark blue and light blue regions represent $t_{\rm Ma}/t_{\rm Ca} < 0.95$ and the orange region indicates the situation of $t_{\rm Ma}/t_{\rm Ca} > 0.95$. For the parameters in (A), (B), (D), (E) and (F), $\gamma_{\rm re}=2.6$. All other parameter setting can be found in SI for details.}
       \label{fig4}
\end{figure}

We found that the magnitude of Marangoni flow can be tuned by adjusting the initial volume fraction ($c_0$) of the solvent with higher surface tension and faster evaporation rate. Fig.\ref{fig4}A shows the time evolution of the magnitude ratio between capillary flow ($v_{\rm Ca}$) and Marangoni flow ($v_{\rm Ma}$) for three deposition patterns at $J_{\rm re}=1.7$. It is clear that both fluid flows are spatially dependent. However, theoretical calculations show that the spatial dependence of the two flows have approximately the same form (see SI Appendix Note S5, S6 in Part 5. Theoretical analysis). Therefore, the ratio of the two flows, $|\kappa|=|v_{\rm Ca}/v_{\rm Ma}|$, is mainly time-dependent with weak spatial dependence. The green line in Fig.\ref{fig4}A indicates that in the case of coffee-ring pattern ($c_0=0.10$), capillary flow is significantly larger than Marangoni flow throughout the evaporation process. Oppositely, for the volcano shape ($c_0=0.90$, purple line), capillary flow is smaller than Marangoni flow. uniform deposition patterns can be achieved only when capillary flow is approximately larger than Marangoni flow, falling within the range $|\kappa|\sim [1,10]$ for $c_0=0.40$, $0.50$, and $0.60$ (orange lines). The radial distribution of the total height-averaged fluid velocity $v_f$ within the droplet at $t=0.5\tau_{\rm ev}$ is theoretically calculated (Fig.~\ref{fig4}B) and experimentally measured using Particle image velocimetry (PIV) (Fig.\ref{fig4}C). For experiments, averaged $v_f$ is measured at the plane $25{\rm \mu m}$ above the substrate at a chosen time $t=0.5\tau_{\rm ev}$ (see Methods for details). Both theoretical and experimental results align well with Fig.\ref{fig4}A: $v_f$ for the coffee-ring (green) is outward from the center to the edge, for the volcano pattern (purple) it is inward from the edge to the center, and for the uniform pattern (orange), it is outward from the center though with a smaller strength compared to the $v_f$ corresponding to the coffee-ring pattern. Furthermore, previous studies indicate that the movement mode of the contact line is important for the deposition patterns \cite{man_2016_ring}. Fig.~\ref{fig4}D presents the time evolution of droplet contact radius (solid lines) and contact angle (dotted lines) of three typical deposited patterns. It is found that the uniform deposition can be obtained only when the contact line contracts slowly within most of the evaporation but rapidly at the end stage, accompanied with a linear decrease of the contact angle ($c_0=0.50$ orange lines).

In addition to the magnitude of the two flows, the duration time of the existence of the two flows is also important in determining the final deposition pattern. The ratio of evaporation rate between the two components, $J_{\rm re}$, can effectively regulate such duration time during which the two flows synergistically convect particles. Fig.\ref{fig4}E shows that when $J_{\rm re}=1.02$ and $5.0$, the value of $|\kappa|$ remains consistently high ($|\kappa|>10$) throughout most of the evaporation process. This is because when $J_{\rm re}$ is small, the concentration gradient along the liquid/vapor interface is small resulting in weak Marangoni flow. On the other hand, when $J_{\rm re}$ is large, the quick loss of the component with faster evaporation rate leads to the transition of a binary droplet to a single component droplet, causing the Marangoni flow to rapidly weaken or even disappear. Consequently, capillary flow dominants in both cases, resulting in a ringlike pattern of deposited particles. In contrast, when $J_{\rm re}$ is $1.7$, $|\kappa|$ stays within the range of $[1,10]$ for most of the evaporation process, indicating a balance in both the magnitude and existence-duration time of capillary and Marangoni flows. Due to the opposing direction of the two flows, a balance is achieved in particle convection, leading to the formation of uniform deposition particles. Fig.\ref{fig4}F presents the theoretical dependence of the ratio of the existing duration time of Marangoni and capillary flows ($t_{\rm Ma}/t_{\rm Ca}$) on $J_{\rm re}$ at $c_0=0.50$, in which the orange region represents the area conducive to achieving uniformly deposited particles. It can be observed that $t_{\rm Ma}/t_{\rm Ca}$ is approximately equal to $1$ ($>0.95$) only within this region, suggesting a suitable range for $J_{\rm re}$ of approximately $[1.1,2.4]$.

\subsection{Adaptability of the method}

\begin{figure}[!tbhp]
       \centering
       \includegraphics[width=0.65\linewidth]{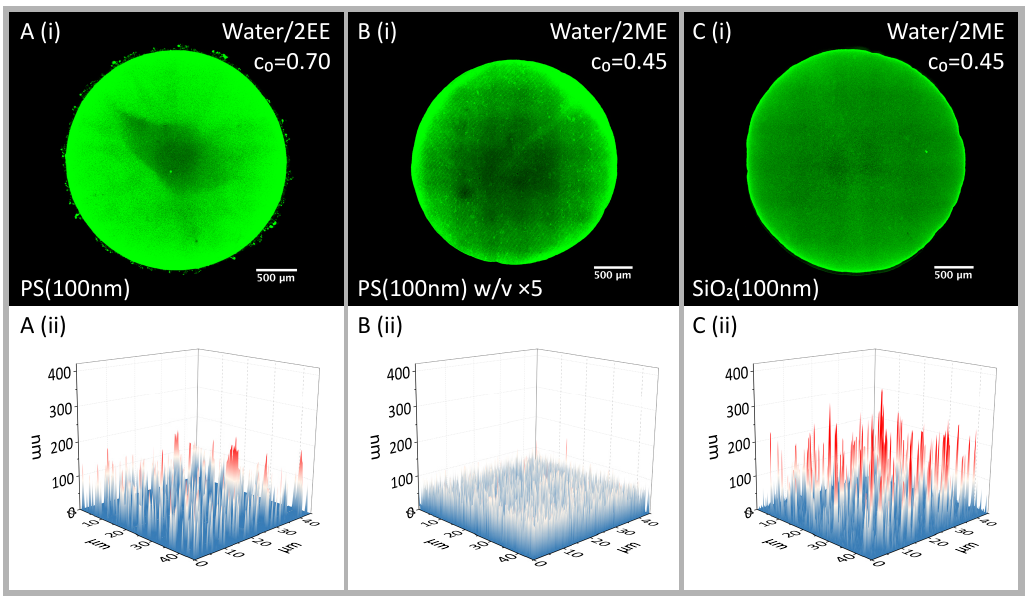}
       \caption{Adaptability of the method. Fluorescence macroscope images and AFM 3D views profiling images of the deposited patterns with different solvent choice (A), different solutes volume concentration vol (B) and different solute particles (C). $c_0$ is the initial volume fraction of water in binary droplets. The color represents the morphology height, same to that in Fig.1E.}
       \label{fig5}
\end{figure}

The synergetic modulation strategy by drying binary droplets allows simple and facile fabrication of uniform pattern of deposition particles, which is readily applicable to various fabrication conditions. Here, we validate the adaptability from three aspects: the solvent choice, the particle concentration, and the particle type. First, we take use of different solvent choice with water/2EE binary droplets. Fig.\ref{fig5}A(i) presents the confocal optical images of deposited particles (PS particles, diameter $D=100nm$). Fig.\ref{fig5}A(ii) is the AFM 3D height distribution of the randomly selected region of Fig.\ref{fig5}A(i), and the height distribution is around $100nm$. This result supports the fact that when the solvent properties are within the range of theoretical parameters metioned above, the uniform deposition can be found by changing the volume ratio of binary solvents, providing more options for solvent selection. Besides, we vary the concentration of the added solute particles to $5\times$ standard concentration, shown in Fig.\ref{fig5}B. Although the light intensity near the contact line $r/R_0=1.0$ is a little higher than other locations, indicating little higher concentration aggregation of solute particles, the morphology in most regions remains uniform. Subsequently, we change the solute types to produce uniform films of different materials. Fig.\ref{fig5}C is the fluorescent images and AFM profiling images of $\rm{SiO}_2$  particles (diameter $D=100nm$) deposites, showing a height distribution around $200nm$. Additionally, we compared this result with the deposition pattern formed by drying pure water droplets ladened with $\rm{SiO}_2$ particles, which exhibit a typical coffee-ring morphology (see SI Appendix Fig.S11). The comparison clearly demonstrates that the drying of binary droplet effectively suppress the CRE observed in pure water droplets, leading to a unifrom deposition pattern.

We also tested quantum dots (QD) particles of a diameter of approximately $D\approx 12nm$. The results showed that the CRE was significantly suppressed (see SI Appendix Fig.S11). However, due to their small size, the diffusion of QD is not solely governed by the flow field. As a result, we cannot obtain uniform depostion pattern of QD. Previous results indeed show that the particle size \cite{Jung_2010_Forces, Wong_2011_Nanochromatography, Weon_2013_Self, Kim_2018_effect, Parthasarathy_2022_experimental, Zolotarev_2022_Monte}, shape \cite{yunker_2011_suppression, Yunker_2012_Influence, Yunker_2013_Effects, Askounis_2015_Effect, Kohri_2019_Ellipsoidal}, and concentration \cite{Kumnorkaew_2009_Effect, Parthasarathy_2021_Further} play a critical role in determing the final deposition patterns. Weon and Je pointed out that smaller nanoparticles are more likely to migrate toward the contact line, facilitating self-pinning more easily than microparticles at the same volume fraction \cite{Weon_2013_Self}. Therefore, further theoretical and experimental investigations are needed to achieve uniform deposition of quantum dots by drying of droplets.

\section{Conclusion}
In summary, the proposed technique demonstrates significant potential as an evaporation-induced method for achieving uniform particle deposition. The Marangoni flow, which is driven by solvent concentration gradients induced by evaporation, provides a distinctive advantage in controlling the deposition patterns over previous studies: precise modulation of flow magnitude can be achieved merely by adjusting the initial volume ratio of the binary mixture, eliminating the need for complex chemical or environmental treatments. Uniform deposition is successfully achieved within the initial water volume ratio range of $c_0=0.40-0.55$ in drying of water/2ME binary droplets, with the root mean square ($\rm Rms$) of the normalized AFM height measurement remaining below $0.1$ ($h(r)/100nm$) and the $\rm Rms$ of the normalized light intensity being less than $0.3$. This method is user-friendly, low-cost, scalable across a range of solute concentrations, additive-free and compatible with various particle types. These advantages pave the way for novel bottom-up printing and coating techniques in physical and materials sciences \cite{Jiao_2023_Mosaic, Yang_2023_StimuliTriggered}.


\section{Experimental Section}
\subsection{Experimental Materials and Preparation}
All chemicals were obtained from commercial sources and used the reagents without any further purification if not otherwise stated. 2-methoxyethanol ($\geq 99.5\%$, Macklin), 2-ethoxyethanol ($\geq 99.5\%$, Macklin), ethylene glycol ($\geq 99.5\%$, Macklin), ethanol ($\geq 99.5\%$, Macklin), isopropyl alcohol ($\geq 99.5\%$, Macklin), fluorescent polystyrene nanospheres (diameter $D=100{\rm nm}$ \& $D=1{\rm \mu m}$, Thermo Fisher), fluorescent silicon dioxide nanospheres (diameter $D=100{\rm nm}$, Zhongkekeyou Co., Ltd), CdSe/ZnS core/shell quantum dots (diameter $D\approx 12{\rm nm}$, Suzhou Mesolight Inc Co., Ltd), silica gel (diameter $D=2-4{\rm mm}$, Macklin), activated charcoal (Macklin) were used as received. 2-methoxyethanol was added to deionized water to configure a solution mixture of water volume fractions ranging from $5\% {\rm vol}$ to $95\% {\rm vol}$. Fluorescent polystyrene nanospheres were added to the solution mixture and followed by ultrasonic dispersion ($300W$) for 5min to make it homogeneous. The typically volume fraction of the nanospheres was $0.0005\% {\rm vol}$. Silicon wafer (Shun Sheng Electronic Technology Co., Ltd) was cleaned sequentially with isopropyl alcohol, deionized water, and ethanol for 5min each, and blown dry with nitrogen. A droplet with a typical volume of $1.5{\rm \mu L}$ was deposited onto a silicon wafer and dried in the humidity cell. The drying process was performed at room temperature. After complete drying of the droplets, the samples were removed from the set-up and further characterized.

\subsection{Experimental Set-up}
To avoid the effect of environmental disturbance, a simple experimental set-up (custom-made humidity cell) was made. The set-up consists of three parts: the exterior is filled with silica gel and activated carbon, which absorbs water and organic matter respectively, to ensure stable vapor pressure during the drying process; the interior is used for placing the silicon wafer; and, a cover is necessary to enclose the entire set-up and reduce interference from the outside environment. All experiments are carried out in environmental conditions with a relative humidity (RH) of $30\%$ to $40\%$ and at room temperature of $20-25^{\circ}{\rm C}$ inside a humidity cell. The RH and temperature are monitored by a sensor (Sensirion SHT85). 

\subsection{Experimental Characterization}
The morphology of thin films was characterized with both fluorescent confocal microscope (Nikon Ti-E) using $10\times/{\rm NA}=0.3$ objective and atomic force microscope (Bruker Dimension FastScan). The contact angle measurement was performed with DataPhysics OCA20 contact angle measuring system. The flow field velocity is characterized by time-lapse fluorescent confocal microscopy (Nikon Ti-E) using $20\times/{\rm NA}=0.5$ objective with the focus plane set above the substrate. During the drying process, a series of images are recorded continuously with an interval time of 30ms for particle image velocimetry (PIV). The PIV measurements are processed by an open-source software based on the publicly available MATLAB code written by W. Thielicke and E. J. Stamhuis \cite{thielicke_2014_pivlabtowards}.

\medskip
\textbf{Supporting Information} \par 
Supporting Information is available from the Wiley Online Library or from the author.

\medskip
\textbf{Acknowledgements} \par 
This work was supported in part by the National Natural Science Foundation of China (NSFC) under Grant No. 22473005, No.21961142020, No.21822302 and No.12072010, the Fundamental Research Funds for the Central University under Grant No.YWF-22-K-101. We also acknowledge the support of the High-Performance Computing Center of Beihang University.

\medskip

\bibliographystyle{MSP}
\bibliography{template}

%

\end{document}